# Manual Character Transmission by Presenting Trajectories of 7mm-high Letters in One Second

Keisuke Hasegawa, Tatsuma Sakurai, Yasutoshi Makino and Hiroyuki Shinoda

*Abstract*— In this paper, we report a method of intuitively transmitting symbolic information to untrained users via only their hands without using any visual or auditory cues. Our simple concept is presenting three-dimensional letter trajectories to the user's hand via a stylus which is mechanically manipulated. By this simple method, in our experiments, participants were able to read 14 mm-high lower-case letters displayed at a rate of one letter per second with an accuracy rate of 71.9% in their first trials, which was improved to 91.3% after a five-minute training period. These results showed small individual differences among participants (standard deviation of 12.7% in the first trials and 6.7% after training). We also found that this accuracy was still retained to a high level (85.1% with SD of 8.2%) even when the letters were reduced to a height of 7 mm. Thus, we revealed that sighted adults potentially possess the ability to read small letters accurately at normal writing speed using their hands.

## I. INTRODUCTION

In our daily lives, it is indispensable for us to access to text information via our sensory systems. Such requirements are met usually with our visual and auditory modalities. On the other hand haptic/kinesthetic modalities are not used unless the case they are the only available ways. Although there are several haptic/kinesthetic techniques of text display used among blind people such as braille (including refreshable braille displays) and Optacon [1], those for sighted people are not in common use. In spite of those techniques being quite useful, they usually require long-term trainings to master, which means that late-blind and sighted people would have greater difficulty in fluently using it. Thus reading by hands are regarded as an esoteric technique only available for well-trained early-blind people.

Many researches refer to the ability of sighted people in manually reading letters. The most straightforward method for conveying alphabets to a reader via hands would be the 'print-on-palm' method, where the reader has letter strokes traced on their palms by a finger, a stylus and so forth [2][3][4]. These studies would fall into 'passive' method for alphabet communications. A different sort of approaches that lets the reader 'actively' read letters touching embossed letters [5][6] with fingers or engraved letters with a stylus[7] are also reported. Unfortunately, these 'passive' and 'active' techniques by and large indicated that an accurate perception of the letters were achieved only when they are considerably large and displayed at least several seconds (and sometimes repeatedly as many times as the readers wanted). In addition, many of

K. H, Y. M and H. S Authors are with the Graduate School of Frontier Sciences, University of Tokyo, 5-1-5, Kashiwanoha, Kashiwa-shi, Chiba-ken, Japan. (e-mail: 'Keisuke_hasegawa, Yasutoshi_makino, Hiroyuki_Shinoda' @k.u-tokyo.ac.jp)

T. S is with TOSHIBA Corporation, Japan.

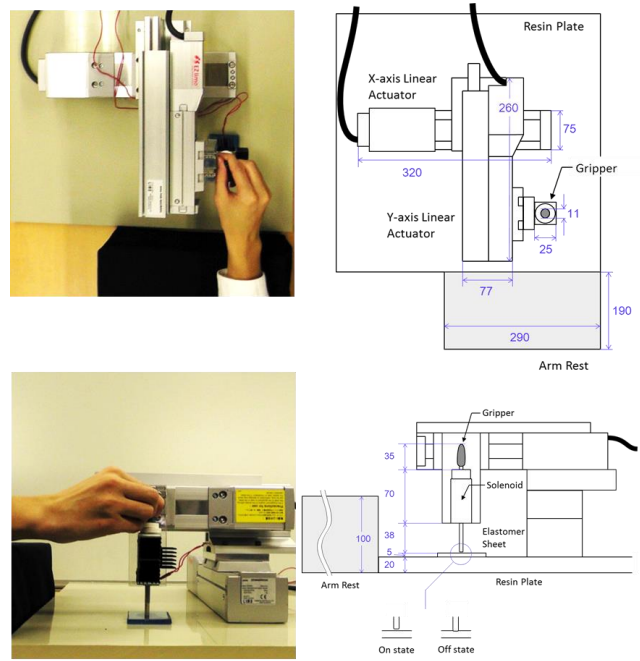

Figure 1. Experimental Device: Top view (top figures) and side view (bottom figures). It is composed of a 2-dimensional linear actuator using stepping motors for the horizontal motion and an on-off state solenoid attached on the actuator for the vertical motion. Participants held the gripper, upper tip of the solenoid covered with rubber.

those studies reported neither on the duration of displaying letters, nor on the reading speeds they accomplished from the perspective of practical information conveyance. Thus, conveying information to sighted people via their hands seems to be regarded as impractical.

For all that, a number of studies focuses on another different methodology, forcing readers to trace trajectories. In [8], experimental participants took part in a task where they traced embossed outline of figures to identify in two ways: one is tracing with fingers moving voluntary and the other is tracing with fingers held and guided along the outline by the experimenter. The performance of the participants was better with the latter condition, when their fingers were guided. This inclination is indicated with another studies where participants held a stylus guided along engraved letter trajectories [7]. With respect to those reports, another studies have found that those performance are possible even when participants have their finger guided along trajectories on a flat plane without any emboss or indentation [8][9]. For practical application, Gesture output [10] succeeds in displaying Graffiti, a single-stroke capital alphabet set via a finger on an electrically manipulated sheet equipped on the surface of tablet computers. Although the exact duration during which each letter was displayed or the letter height were not described, Trained participants were able to accurately read tens-of millimeter-high letters displayed for several seconds.

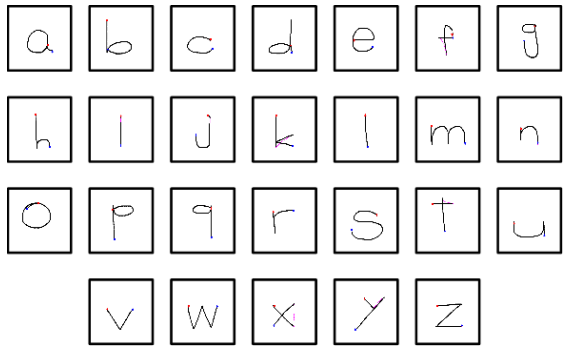

Figure 2. Displayed letter trajectories. Red points indicate the starting point in each trajectory terminated at blue points. The stylus tip is supposed to be on the paper along the bald black lines and to be lifted up over fine purple lines. All letters are depicted in the same scaling.

While Gesture output puts the focus rather on offering two users mutual tactual communications, we expect the possibility of this 'forced writing' method in efficiently transmitting text information via hands. We anticipated that the 'forced writing' to be more intuitive method than the 'print-on-palm' or fumbling embossed or engraved letters because the sighted and late-blind people have background experiences of repeatedly writing letters so many times.

In the work reported in this paper, we experimentally demonstrated that sighted people can accurately read lower-case alphabets with their hand forced to trace letter trajectories via an electronically manipulated stylus they held, whose movement is as fast (one second or even half a second with degraded accuracy) and large (14 or 7 mm in height) as the way we write actual letters with a pen. Experiments using an entire set of 26 alphabets showed that the accurate recognition rate of 91.3% for 14mm-high letters and 85.1% for 7-mm high letters in a pace of one letter per second. This performance were achieved after each of the participants had taken only 5-minute trainings. These results indicate that a much faster and smaller movement than what is conventionally believed to be necessary is sufficient for transmitting alphabets to users via hands, with a considerably short training period. For the experiments, we fabricated a device consisted of two simple components: two-dimensional linear actuators and an on/off two-way solenoid attached to them, which simulates the movement of a pen tracing letter trajectories. The two-dimensional linear actuators were driven to exert force strong enough to force the participants' hands to trace pre-programmed letter trajectories with a great repeatability unaffected by the gripping force of the participants.

Several studies have shown that there are some cerebral link between writing motion and alphabet perception [12]. In the field of teaching, researches have pointed out that kinesthetic feedback would lead to a more efficient learning of hand writing [13][14][15]. There are also other types of studies on tactile/kinesthetic-auditory-visual interaction, such as activation of the pre-motor area by visual reading of hand-written words [16] or auditory-tactile coupling in comprehension of uttered syllables accompanied by tactile stimulation [17]. Thus, correlation between tactile/kinesthetic stimuli and text recognition has been implied in cognitive and cerebral investigations. It is true these issues are pretty interesting to address, we do not associate our results with them directly since much more experimental evidence would be necessary for doing so. We simply report on the performances in our experiments and evaluate them from the perspective of practical symbolic information conveyance.

Note that what we describe in the following part of the paper does not necessarily aim to the conventional 'sensory substitution' compensating lost modalities with another ones.It is expected that a technique which allows us to manually read texts would profit not only visually-impaired people but also sighted people. Since our method employs past experience in writing letters, it would only be intuitive for late-blind or sighted people. Nevertheless, our contribution would create a new practical path to tactual reading available not only for those who have engaged themselves in such long-term trainings as required in braille.

## II. CHARACTER DISPLAY SYSTEM

Figure 1 shows a schematic diagram of the system. An on/off two-way solenoid actuator, which moves vertically, was attached to a two-dimensional horizontally moving stage. Horizontal motions represented the trajectories of letters, whereas vertical motions imitated the contact of the tip of a pen with paper (in the actual experiment a plastic plate was placed under the solenoid instead of paper). We used a pair of linear actuators driven with stepping motors (models EZS3D005-A and EZS6D005-A, Oriental Motor Co., Ltd.) joined to each other perpendicularly. The workspace of it was 50 mm x 50 mm and its spatial resolution is 0.01 mm. They were mounted on the plastic plate. The on/off two-way solenoid (model 8.M14-02-62-12 VCC-100\%, Mecalectro) was attached on the stage of the linear actuators with its lower tip pointing towards the plastic plate. The vertical movement of the tip was 1 mm. The linear actuators and the solenoid were electronically controlled by the computer via a microcontroller board (Arduino Mega 2560, product of Arduino), resulting in three-dimensional movement of the tip of the solenoid. On the lower side of the solenoid, metal spacers were attached in order to adjust the vertical movement range of the gripper. During the experiments, participants held the gripper, at the upper side of the solenoid stylus, with their arms placed on an arm rest, during vertical up/down movement of the stylus and lateral movement along the letter trajectory. We placed an elastomer sheet under the stylus in order to attenuate the sound caused when the stylus hit the surface.

Trajectories of lower-case letters were recorded from actual letters written by writers on a graphics tablet (model PTH-850KO, product of WACOM). The extracted trajectories are shown in Figure 2. The letters were drawn with no serifs. The movement of the stylus tip, including the up/down contact state between the pen and paper, was also recorded. The recorded trajectories were temporally smoothened by FIR filtering for removing jumpy horizontal movements of the stylus. The character shapes were appropriately deformed from the standard typefaces for enhancing the differences among confusing letters (such as the vertical bar excessively prolonged in 'd' to be differentiated from 'a'). This designing was found to be necessary for a better performance in previously conducted preliminary experiments showing that standard typefaces or even the reproduced strokes of the participants' own would yield more confusions. In the experiments, spatially resized trajectories were presented to the participants according to the experimental conditions. They were also temporally stretched so that all letters were displayed within a constant duration. During these processes, temporal changes of trajectory velocities were also scaled to maintain similarity.

III. Experiments

A. Methods and Procedures

We conducted two experiments in which participants were asked to identify the presented letters. Participants were told in advance that the presented letters were randomly chosen from 26 lower-case letters. For each trial a letter was displayed only once. Participants answered verbally what letter was displayed. No equivocal answer was permitted. Throughout the experiments, participants wore headsets playing white noise to nullify auditory cues caused by the mechanical movement of the experimental system. During the tasks the whole device was covered with curtains to nullify any visual cues. Participants were instructed to concentrate on tactile and kinesthetic sensations and to keep their view away from their arms. There was no time limit for the participants' responses. The letters were displayed one by one, intermitted by the participants' utterance. All participants were naive: they had never experienced the experimental system before.

We tested 20 naive participants, including 8 female and 12 male. Their ages ranged from 23 to 53. One participant was Chinese and the others were all Japanese. All participants were right-handed. Each of them was rewarded with a 500 JPY pre-paid card for purchasing books. Prior to the first experiment, participants adjusted their arm postures while the tip of the stylus went up and down, so that they learned to apply appropriate load on the stylus. This procedure was necessary to make sure that the solenoid hit the plastic plate when it was in the 'off' state. In all experimental sets, each letter was displayed twice in a random sequence, totaling 52 trials in all. Participants did not know the number of trials in every experimental set. The following two experiments were conducted with all of the participants in the same order (the first followed by the second)

B. First Trial for Naive Participants

At the beginning part of the first experiment, we demonstrated how the system worked. The participants could see the movement of the stylus for every letter, one time for each letter. During this demonstration, participants were allowed to see the complete set of the letters displayed on a board, as seen in Figure 2, but were not permitted to touch the device. After the demonstration, the sample alphabet board was removed, and the participants worked on the identification task. They identified letters whose average height was set to 14 mm at a pace of one second per letter. It took 5 minutes at the longest for the participants to finish this first trial.

C. Second Trials after a Five-minute Training Period

In the second experiment, a short five-minute training period was executed before the identification tasks. We thought that the stroke writing styles would be different among participants (for example, some of them would start writing 't' with the horizontal bar first, whereas others would write the vertical bar first), and therefore, the training was performed with the aim of getting participants accustomed to the presented letter strokes. In the training the participants guessed displayed letters with an average height of 14 mm at a rate of one second per letter. Each letter was chosen by the experimenters, and the participants were informed whether they were correct for every guess. This procedure lasted for five minutes for each participant. After this five-minute training period, first, participants performed the identification task under the same experimental conditions as those used in the first trial (14mm/1000ms). Subsequently, they performed the identification task in three other sets (7mm/1000ms, 14mm/500ms, and 7mm/500ms) with a randomized order.

D. Results

The first experiment showed that naive participants were able to identify displayed letters with an average accuracy rate of 71.9% (standard deviation, SD=11.3%) when the average letter height was 14 mm and the duration was 1000 ms. The second experiment showed the effect of the five-minute training period. The accuracy rate was improved to 91.3% (SD=6.7%) (Figure 3). A paired t-test showed that the five-minute training period resulted in a significant ($p < 10^{-8}$) difference.

The average accuracy rates in the second trials were, 91.3% (SD = 6.7%) for the (14 mm/1000 ms) condition as stated above, 85.1% (SD = 8.2%) for (7 mm/1000 ms), 72.5% (SD = 11.12%) for (14 mm/500 ms), and 60.8% (SD = 12.5%) for (7 mm/500 ms), respectively. The effect of reducing the size of the letters was less evident than that of increasing the writing speed (Figure 4).

By applying two-way analysis of variance, we verified that both the letter height ($p < 10^{-4}$) and writing speed ($p < 10^{-15}$) had significant effects on the accuracy rate, whereas a combination of the two did not ($p > 0.05$). Under all five experimental conditions (first trial and four experiments after 5-minute trial), it was concluded that there was no significant difference between male and female participants from the results of a t-test ($p > 0.05$).

Confusion matrices of pooled participants' response and actually displayed alphabets for the first and second trials is shown in (Table 1).

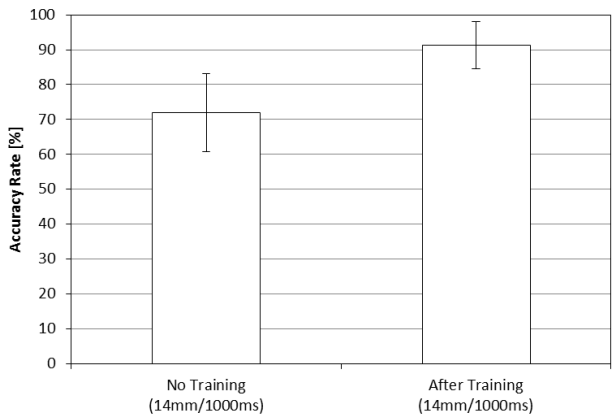

Figure 3. Accuracy rate graphs with standard deviation bars under the five experimental conduitions.

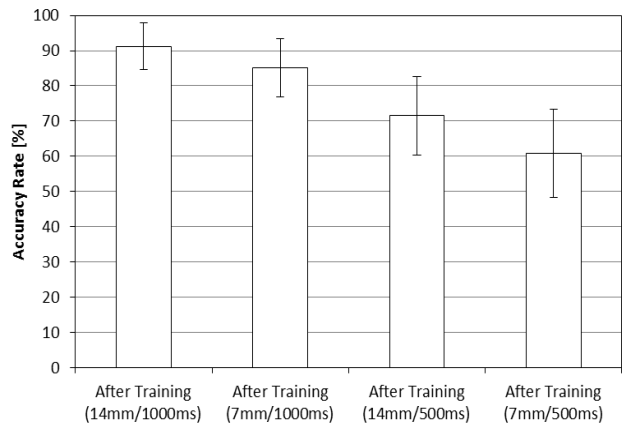

Figure 4. Accuracy rate graphs with standard deviation bars under the four experiental conditions.

IV. DISCUSSIONS

The effect of the training was significant, which boosted the accuracy rate by nearly 20 percentage points from the first trials. The fact that none of the participants could reach an accuracy rate of 90\% at the initial trial reflects the necessity of the trainings for getting accustomed to the displayed letter set, since personalized letter strokes have much difference among individuals [18].

It is also notable that the accuracy rate in reading 7 mm-high letters was 85.1% (SD=8.2%), which was only 6 percentage points less than the accuracy rate with 14 mm-high letters. The ability to read 7 mm-high letters with the movement of hand joints is consistent with the cognitive detection threshold of joint movement verified in previous studies [19] (The slowest stroke in our experiment was 7 mm/s, which corresponds to 1.15 deg/s in horizontal elbow movement for an arm measuring 35 cm from the elbow to the fingertips). This small workspace, occupying a 7 mm square, implies that the device can be as small as a single key of a standard keyboard, theoretically. On the other hand, the effect of shortening the display period was more critical in deteriorating the accuracy rates.

In Table 1, some specific pairs of letters are seen to be more confusing than the others. Many participants pointed out that some of the letters were indiscernible in their first trials, as the displayed strokes were different from their habitual ones. This inclination is seen such pairs as 't-f', 's-g' and 'd-a' (all indicated in 'displayed-response' manners). Those confusion was corrected after the participants had learned the 'right' strokes in the designed letter set. Yet the difficulties in differentiating some letters from others, namely 'x-y', 'e-c' and 'w-n' confusion, are seen to remain even after the trainings. These are because of their similar trajectories, which we expect that a suitable design of stroke patterns and longer training can improve the recognition rate.

The achievable reading speeds corresponding to our experiments are speculated to be 60 letters per minute (with a rate of 1000 ms/letter) and 120 letters per minute (with a rate of 500 ms/letter), excluding the time for giving answers. The former speed, which guarantees accurate symbolic communication, is almost half of the typical adult handwriting speed of 130 letters per minute [20]. We should state that the actual reading speed would certainly be different from the speculated number above since there are a great difference between perceiving each letter one by one and comprehending a sequence of letters as a word. An optimistic factor for expecting better actual reading pace than experimentally obtained is that we anticipate and speculate in comprehending words as we see and hear each letter of them. It is the case that sometimes we can identify a word before knowing every letter of it thanks to the context and dictionaries in our brains. Additionally, if we allow the use of trajectory patterns designed specifically for hand-reading, like shorthand, the reading pace might be improved.

V. CONSLUSION

We have experimentally demonstrated that sighted people are able to read small letters using only their hands at normal writing speed by presenting 3D trajectories of standard-sized written letters, including up-and-down motion. With no preliminary training, the accuracy rate for reading 14 mm-high letters was 71.9%, which improved to 91.3% after only a five-minute training period. In these results, little individual variations were seen (SD = 12.7% in the first trials and 6.7% after training), which suggests that the method we employed here is generally effective among sighted adults. When the letters were downsized to 7 mm in height, the accuracy rate remained 85.1% (SD = 8.2%).

A significant contribution of this research is that it opens up a new field of human--computer interfaces that go beyond the conventional limitations of visual and auditory modalities. We expect that the communication rate of our method will be much improved by practice and technical modification of the interface design. The method can be embedded in various devices in principle.

The short-period trainings executed in our experiments can be operated with only tactile feedback to let the trainees know whether their guess about presented alphabets are correct. Therefore, it is possible for acquired deafblind people who had written alphabets before they lost their sights to read by our method with much shorter trainings compared to those necessary for mastering other tactile information transmission methods such as braille.

Table 1. Pooled confusion matrices: tables (A), (B) correspond to the experimental conditions denoted below the tables. Numbers in each cell indicate the counts of participants' answers according to combinations of displayed letters and participants' responses. The diagonal components correspond to correct answers. There should be 40 correct answers at most (2 answers from each of 20 participants).

There is another positive aspect of our method: since symbolic information systems which entirely rely on tactile/kinaesthetic sensations are intrinsically invisible and inaudible, our method can be applied to such secure information displays as would be used for showing the PIN of newly opened bank accounts to the customers at bank windows.

This research is the first step to realize manual alphabetic information display. As stated in the discussions, letter identification and word comprehension is considerably different. Our next challenge would be an interactive manual text display system which allows users to back and skip to the next word at any time.

As demonstrated in Gesture output, there is a possibility for applying our method in portable usage, although there seems to be many problems to be solved in advance: identification of appropriate hand postures, establishment of sufficiently strong and small mechanical structures, and a good substitution for up-and motion of the method.

ACKNOWLEDGMENT

This work was partly supported by JST CREST and JSPS KAKENHI Grant Number 25240032.